 \documentclass[12pt]{iopart}
 \usepackage{epsfig}
 \newcommand{\sqrtsNN}{\mbox{$\sqrt{\mathrm{s}_{_{\mathrm{NN}}}}$}}
 \newcommand{\pT}{$p_{T}$}
 \newcommand{\Kzs}{$K^{0}_{s}$}
 \newcommand{\Kzsm}{K^{0}_{s}}
 \newcommand{\RdA}{$R_{dA}$}
 \newcommand{\RdAu}{$R_{dAu}$}
 \newcommand{\RCP}{$R^{dAu}_{CP}$}
 
\begin{document}
 \jl{4}
\title[STAR d+Au Identified hadrons]{Identified Particle Dependence of Nuclear Modification Factors in d+Au Collisions at RHIC}

 \author{L S Barnby\dag for the STAR Collaboration\footnote[2]{For the full author list and acknowledgements, see Appendix ``Collaborations'' of this volume.}}
 
 \address{\dag\ School  of Physics and Astronomy, University of Birmingham, Edgbaston, Birmingham B15 2TT, UK}
\ead{L.Barnby@bham.ac.uk}

 \begin{abstract}
 We present transverse momentum spectra from d+Au collisions at \sqrtsNN~=~200 GeV for identified hadrons at mid-rapidity and unidentified hadrons in the forward rapidity region using the Solenoidal Tracker at RHIC. We scale these spectra by the mean number of binary collisions to form a transverse momentum dependent ratio to spectra from p+p collisions, showing that the Cronin effect measured at lower centre-of-mass energies is also present at this higher energy. The data also suggest a dependence of the Cronin effect on the hadron species. We also compare central d+Au collisions to more peripheral ones, contrasting the behaviour to that observed in Au+Au collisions and remark on unidentified hadrons at forward rapidities which show a qualitatively different behaviour.
\end{abstract}


\section{Introduction}
The addition of d+Au collisions to the experimental programme at the Relativistic Heavy-Ion Collider (RHIC) has allowed an important control system to be studied and compared to Au+Au collisions. Recent Au+Au observations made at RHIC include high-\pT\ suppression \cite{STARhiptsup} and the related disappearance of back-to-back correlations \cite{STARAuAuBtoB}. Performing the same analyses with d+Au can distinguish new phenomena related to effects of the final state rather than to the initial state of the Au nucleus, as a strongly interacting final state should not form in d+Au collisions. It was quickly established that high-\pT\ suppression is not present in d+Au collisions, nor are back-to-back correlations suppressed \cite{STARdAuBtoB}. Another key observation is the difference in nuclear modification factors for various hadronic species in the intermediate \pT\ region (2-5 GeV/c) \cite{STARK0LamRcp}.  These measurements sparked renewed interest in previous p+N collision experiments at lower energies, where a \pT-dependent effect on hadron production, with an enhancement in the \pT\ region above 1 GeV/c, was discovered \cite{CroninInitial}. This enhancement, termed the ``Cronin effect'', shows differences between protons, pions and kaons. The measurements for pions, which are the most precise, showed the magnitude of the effect decreasing with increasing centre-of-mass energy, $\sqrt{s} = 20-40$ GeV \cite{StraubIDCronin}. By investigating identified hadron production at $\sqrt{s} = 200$ GeV we can see whether the difference in nuclear modification factors with hadron species in Au+Au collisions could be related to a difference in the Cronin effect. Looking at forward rapidities, where the saturation scale is larger, we may be sensitive to effects predicted by saturation models~\cite{CGCPre}.

\section{Experiment} 
In this analysis the  following Solenoidal Tracker at RHIC (STAR) tracking detectors were used~\cite{NIM_STAR}; a large cylindrical Time Projection Chamber (TPC) at mid-rapidity~\cite{NIM_TPC}, further TPCs in the forward regions (FTPC), around rapidity, $y\approx\pm3$~\cite{NIM_FTPC}. Zero-Degree Calorimeters (ZDC) which detect spectator neutrons near to beam rapidity are available for triggering~\cite{NIM_ZDC}. A minimum bias trigger consisting of a coincident ZDC signal from spectator neutrons on the Au side and a timing signal indicating the window for the bunch crossing was employed to collect around 20 million events. In the offline analysis these were reduced to 10 million events by insisting that the collision occurred within $\pm 50$ cm of the centre of the TPC ensuring that reliable efficiency and acceptance calculations could be made. Particle identification relied on a time-of-flight (TOF) detector which covers the rapidity interval $-0.5 \le y \le 0.0$~\cite{NIM_TOF}, where the negative rapidity represents the Au-side.  Additionally, weakly decaying neutral strange particles were identified using the TPC by their charged daughter decay modes ($\Lambda \rightarrow p\pi^{-}, \Kzsm \rightarrow \pi^{+}\pi^{-}$)~\cite{LamontThesis} and kaons by their 'kink' decay inside the TPC gas volume ($K^{\pm} \rightarrow \mu^{\pm}\nu,K^{\pm} \rightarrow \pi^{\pm}\pi^{0}$)~\cite{DengThesis}.

\begin{figure}
\begin{center}
	\begin{minipage}[c]{0.58\textwidth}
		\epsfig{file=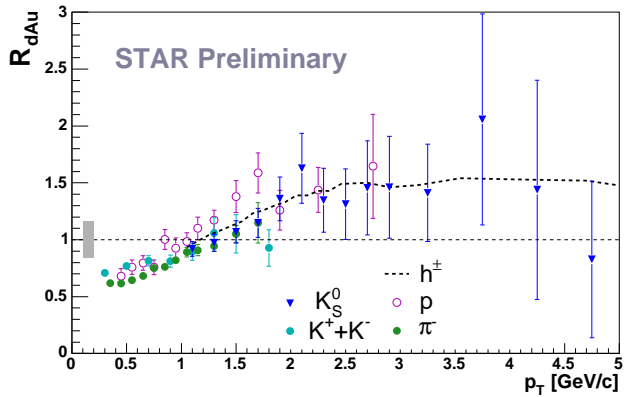,width=\linewidth}
	\end{minipage}
	\hspace{-0.1\textwidth}
	\begin{minipage}[c]{0.48\textwidth}
		\caption{Nuclear modification factors at mid-rapidity as a function of \pT\ for kaons with protons and pions from the TOF detector and unidentified charged hadrons shown for comparison.\label{Fig_RdAuId}}
	\end{minipage}\
\end{center}
\end{figure}

\begin{figure}
\begin{center}
	\begin{minipage}[c]{0.58\textwidth}
		\epsfig{file=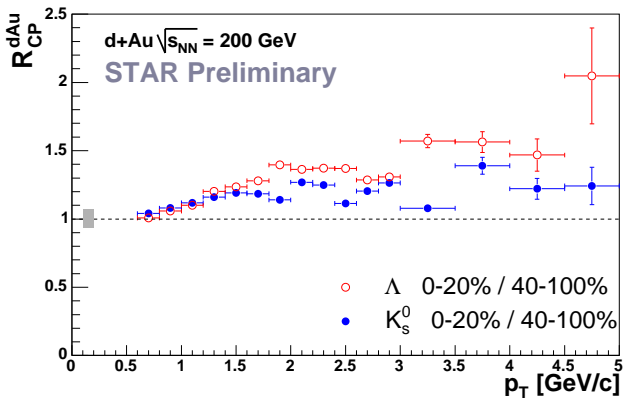,width=\linewidth}
	\end{minipage}
	\hspace{-0.1\textwidth}
	\begin{minipage}[c]{0.48\textwidth}
		\caption{Ratio of the \pT\ spectra, scaled by the mean number of binary collisions, from central collisions (0-20$\%$) to more peripheral collisions (40-100$\%$) for $\Lambda$ and \Kzs. The grey box indicates the scaling uncertainty.\label{Fig_RCPLamK0}}
	\end{minipage}\
\end{center}
\end{figure}

\section{Results}
\begin{figure}
\begin{center}
	\begin{minipage}[c]{0.58\textwidth}
		\epsfig{file=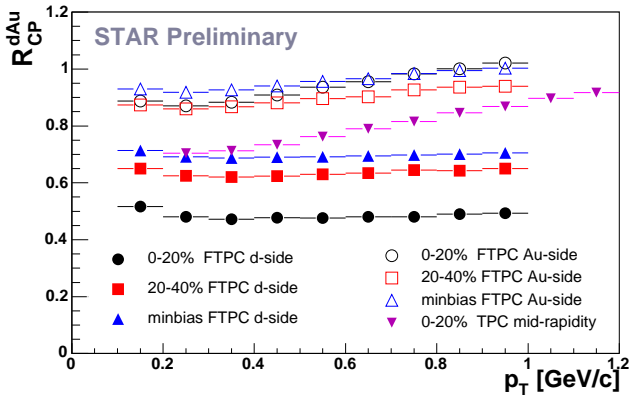,width=\linewidth}
	\end{minipage}
	\hspace{-0.1\textwidth}
	\begin{minipage}[c]{0.48\textwidth}
		\caption{The ratio of the production of charged hadrons to their production in the most peripheral centrality bin as a function of \pT\ for various centrality classes for forward rapidities on the d- and Au-going sides.\label{Fig_ForwardRcp}}
	\end{minipage}\
\end{center}
\end{figure}

Neutral strange particle \pT\ spectra were obtained at mid-rapidity ($|y| < 0.5$) up to 5 GeV/c and up to 2 GeV/c for charged kaons. Spectra from the TOF up to \pT\ of 1.6 and 3 GeV/c for pions and protons respectively were measured~\cite{STARTOF}. Although the FTPCs have some acceptance for pseudorapidity $\eta$ throughout the range 2.5 to 4.0, the measurement was restricted to the interval $3.0 < \eta < 3.2$, where the acceptance is good for a large range in \pT, which avoids problems when extrapolating the spectra. The \RdA\ calculation also requires spectra from p+p collisions and those for $\Lambda$\ and \Kzs\ are presented elsewhere \cite{STARstrangepp}. The nuclear modification factor, \RdAu\, used to look at the nuclear effects on hadron production in d+Au, is calculated as the ratio to the production in p+p with a scaling by the number of binary collisions
\begin{equation*}
R_{dAu} = \frac{1/p_T \cdot d^{2}N/dp_T dy}{\langle N_{bin}\rangle /\sigma^{pp}_{inel}d^{2} \cdot 1/p_T \cdot \sigma^{pp}_{inel}/dp_T dy}
\end{equation*}
where $\langle N_{bin} \rangle$ is the mean number of binary collisions, calculated in a Glauber model, and has a value of 7.5 for minimum bias d+Au collisions~\cite{STARdAuBtoB}. \Fref{Fig_RdAuId} shows the result for \Kzs\ and averaged $K^{+}$ and $K^{-}$, along with previously reported results for protons, pions~\cite{STARTOF} and unidentified negative hadrons~\cite{STARdAuBtoB}. The uncertainty in the scaling is represented by the grey box at one. Unfortunately the corrections needed to produce the $\Lambda$ result were not complete at the time of the presentation. However, it is possible to access information about the nuclear modification factor by forming a ratio of central to peripheral d+Au collisions, \RCP\, similar to \RdAu. Centrality classes for the analysis of mid-rapidity spectra are defined by cutting on the multiplicity observed in the FTPC on the Au side. The measured distribution was shown to match well with the expectation from a Glauber model \cite{STARdAuBtoB}. Centrality classes corresponding to the most central $20\% $ and least central $60\%$ were defined with $\langle N_{bin}\rangle$ of 15.0 and 4.0 respectively. This ratio for $\Lambda$ and \Kzs\ is shown in \fref{Fig_RCPLamK0}. Efforts to use the FTPC for identifying strange particles are still at an early stage \cite{STARforwardstrange} but \RCP\ can be formed in a limited \pT\ range for charged hadron spectra for both the d and Au sides as shown in \fref{Fig_ForwardRcp}. In this case the centrality definition uses the multiplicity at mid-rapidity as the estimator. 

\section{Discussion}
The identified \RdAu\ results show that a significant Cronin enhancement is still present at this centre-of-mass energy and for \Kzs\ at higher \pT\ is consistent with the value found in $\sqrt{s} = 40$ GeV p+W collisions~\cite{StraubIDCronin}. The values for protons are systematically above those for \Kzs\  at low \pT\ but then become indistinguishable at higher \pT\ in part due to the increasing statistical uncertainties. \RCP\ values for $\Lambda$ and \Kzs\ seem to show that the nuclear modification for $\Lambda$ is consistently higher over the full measured range in \pT\ in agreement with other results \cite{PHENIXidRdA}. It is not really possible to generalise and say whether this effect is mass-related or a meson-baryon ordering effect without measuring further hadron species to higher \pT. The size of the effect is rather small in relation to the large differences seen in the nuclear modification factors when comparing central and peripheral Au+Au collisions. This indicates that the Au+Au system is exhibiting an effect due to the final state formed in the collision. A more dramatic centrality dependent effect in d+Au is seen at forward rapidities. Here, there is a {\em suppression} with increasing centrality  on the d-side but almost no difference on the Au-side.


\end{document}